\documentclass[12pt]{article}

\usepackage[superscript]{cite}
\usepackage{amsmath}
\usepackage{braket}
\usepackage{soul}

\usepackage{times}

\usepackage{graphicx}
\graphicspath{ {./images/} }

\newcounter{lastnote}

\title{Direct measurement of key exciton properties: energy, dynamics and spatial distribution of the wave function}

\author
{Shuo Dong,$^{1,\dagger,\ast}$ Michele Puppin,$^{1,2,\dagger}$ Tommaso Pincelli,$^{1}$\\
Samuel Beaulieu,$^{1}$Dominik Christiansen,$^{3}$
Hannes Hübener,$^{4}$\\
Christopher W. Nicholson,$^{1,5}$
R. Patrick Xian,$^{1}$
Maciej Dendzik,$^{1,6}$\\
Yunpei Deng, $^{1,7}$
Yoav William Windsor, $^{1}$
Malte Selig,$^{3}$\\
Ermin Malic,$^{8}$
Angel Rubio,$^{4}$
Andreas Knorr,$^{3}$\\
Martin Wolf,$^{1}$
Laurenz Rettig,$^{1\ast}$
Ralph Ernstorfer$^{1\ast}$\\
\\
\normalsize{$^{1}$Fritz-Haber-Institut der Max-Planck-Gesellschaft, Faradayweg 4-6, 14195 Berlin, Germany}\\
\normalsize{$^{2}$Laboratoire de Spectroscopie Ultrarapide and Lausanne Centre for Ultrafast Science (LACUS),}\\
\normalsize{École polytechnique fédérale de Lausanne, ISIC, Station 6, CH-1015 Lausanne, Switzerland}\\
\normalsize{$^{3}$Institut für Theoretische Physik, Nichtlineare Optik und Quantenelektronik,}\\
\normalsize{Technische Universität Berlin, 10623 Berlin, Germany}\\
\normalsize{$^{4}$Max Planck Institute for the Structure and Dynamics of Matter and}\\
\normalsize{Center for Free Electron Laser Science, Luruper Chaussee 149, 22761 Hamburg, Germany}\\
\normalsize{$^{5}$Département de Physique and Fribourg Center for Nanomaterials,}\\
\normalsize{Université de Fribourg, CH-1700 Fribourg, Switzerland}\\
\normalsize{$^{6}$Department of Applied Physics, KTH Royal Institute of Technology, }\\
\normalsize{Hannes Alfvéns väg 12, 114 19 Stockholm, Sweden}\\
\normalsize{$^{7}$SwissFEL, Paul Scherrer Institute, Villigen, Switzerland}\\
\normalsize{$^{8}$Chalmers University of Technology, Department of Physics, 412 96 Gothenburg, Sweden}\\
\\
\normalsize{$\dagger$ These authors contributed equally to this work}\\
\normalsize{$^\ast$To whom correspondence should be addressed;}\\
\normalsize{E-mail: dong@fhi-berlin.mpg.de, rettig@fhi-berlin.mpg.de, ernstorfer@fhi-berlin.mpg.de.}
}

\date{}

\begin{document} 


\baselineskip24pt


\maketitle 

\newpage

\begin{center}
\ul{\textbf{Abstract} }
\end{center}

\noindent
Excitons, Coulomb-bound electron-hole pairs, are the fundamental excitations governing the optoelectronic properties of semiconductors. 
While optical signatures of excitons have been studied extensively, experimental access to the excitonic wave function itself has been elusive.
Using multidimensional photoemission spectroscopy, we present a momentum-, energy- and time-resolved perspective on excitons in the layered semiconductor WSe$_2$. 
By tuning the excitation wavelength, we determine the energy-momentum signature of bright exciton formation and its difference from conventional single-particle excited states.
The multidimensional data allows to retrieve fundamental exciton properties like the binding energy and the exciton-lattice coupling and to reconstruct the real-space excitonic distribution function via Fourier transform. 
All quantities are in excellent agreement with microscopic calculations. Our approach provides a full characterization of the exciton properties and is applicable to bright and dark excitons in semiconducting materials, heterostructures and devices. 
\vspace{1cm}

\noindent
Excitons, bound electron-hole quasi-particles carrying energy and momentum but no net charge, are fundamental excitations of semiconductors and insulators arising from light-matter interaction\cite{frenkel1931transformation}. 
An initial excitonic polarization induced by a light field (often referred to as coherent excitons and e.g.~detected by optical absorption spectroscopy) rapidly loses coherence with the driving field and dephases into a population of incoherent excitonic states\cite{perfetto2016first,christiansen2019theory}. 
The generated excitons propagate in solid-state materials through diffusion\cite{wang2018electrical,kaviraj2019physics} and eventually release their energy e.g.~in the form of luminescence (photon), lattice excitation (phonon) or dissociation into single charged quasi-particles\cite{palummo2015exciton,yuan2017exciton,christiansen2017phonon,steinhoff2017exciton}. 
Understanding exciton physics is of capital importance for advanced photonic and optoelectronic applications including photovoltaics. 
Layered transition metal dichalcogenide (TMDC) semiconductors exhibit rich exciton physics even at room temperature due to strong Coulomb interaction\cite{wang2018colloquium}. 
Excitons in TMDCs feature large oscillator strength\cite{wang2015giant} and their inter- and intra-band dynamics have been extensively investigated\cite{li2014measurement,pollmann2015resonant,selig2016excitonic}. 
Moreover, strong spin-orbit coupling and broken inversion symmetry in each crystalline trilayer lead to a locking between spin, valley and layer degrees of freedom, which started a surge of valley physics studies \cite{zeng2012valley,mak2012control,bertoni2016generation}.

A large portion of the research on excitonic phenomena in TMDCs adopts optical spectroscopic techniques\cite{wang2018colloquium,li2014measurement,selig2016excitonic,zeng2012valley,mak2012control,park2018direct,yao2017optically,klein2019impact,chernikov2014exciton}, which only access bright excitonic transitions with near-zero momentum transfer. While techniques such as time-resolved THz spectroscopy also allow probing optically dark excitons via internal quantum transitions\cite{pollmann2015resonant}, finite-momentum excitons which lie outside the radiative light cone remain inaccessible to such methods.
This limitation is overcome by time- and angle-resolved photoemission spectroscopy (trARPES), a spectroscopic tool accessing excited states, including excitons, in energy-momentum space and on ultrafast timescales\cite{bertoni2016generation, puppin2019time, weinelt2004dynamics, madeo2020directly}.
Here, we reveal the characteristics of the excitonic wave function in the photoemission signal of the prototypical layered TMDC semiconductor 2\emph{H}-WSe$_2$ and establish that all fundamental exciton properties are encoded in the trARPES signal's energy, time, and momentum dimensions: the exciton binding energy, its self-energy as a measure of the exciton-lattice coupling, as well as the real-space distribution of the excitonic wave function. 

\begin{figure}
\begin{center}
\includegraphics[width=16cm,keepaspectratio=true]{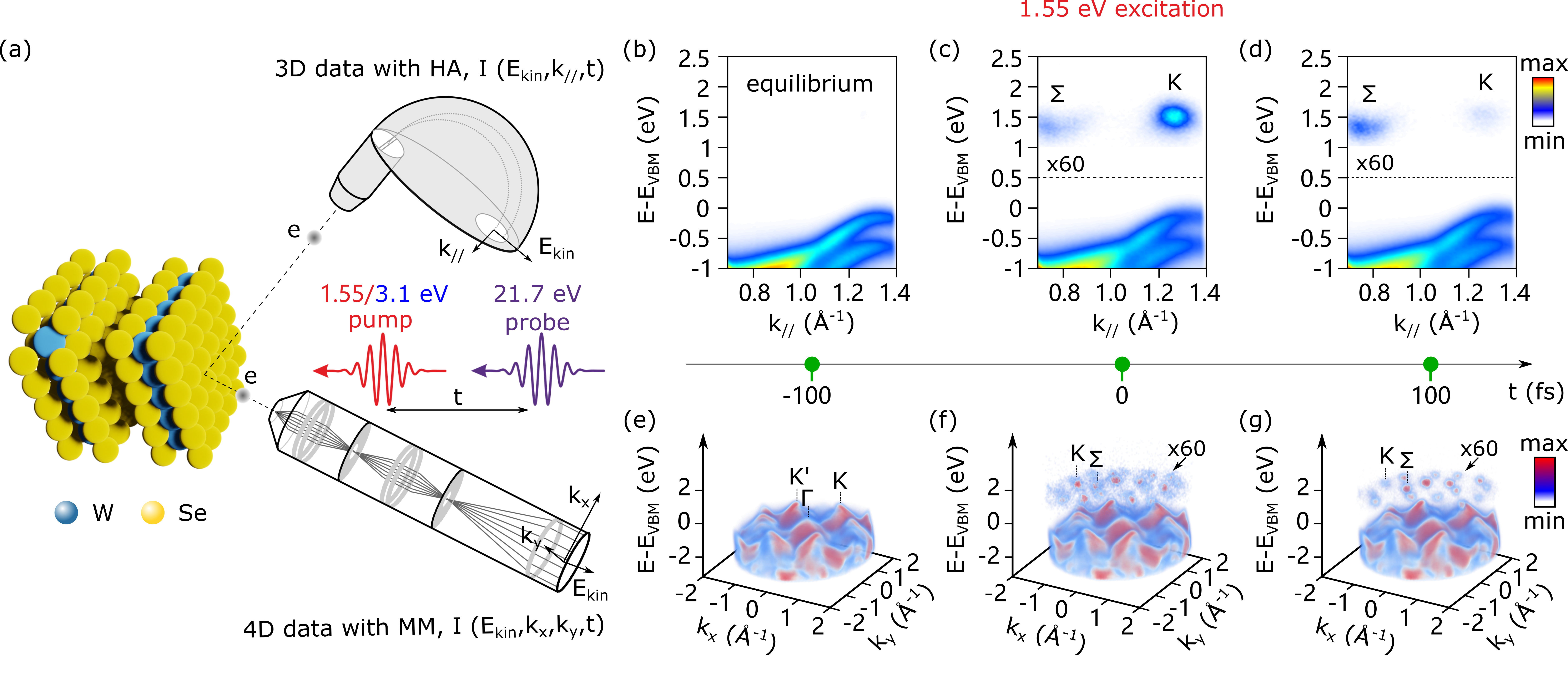}
\caption{\textbf{Multidimensional photoemission spectroscopy of excitons in $\mathbf{WSe_2}$}: (a) Using near-infrared (1.55 eV) or UV (3.1 eV) pump and XUV probe (21.7 eV), we performed trARPES measurements in bulk WSe$_2$ with two types of photoelectron detectors: hemispherical analyser (HA) or time-of-flight momentum microscope (MM). (b) With the HA, the equilibrium band structure in a finite momentum window is found at negative pump-probe delay, $t=-100$ fs, showing the spin-orbit split valence bands near the K-points. Upon 1.55 eV excitation, the excited state dynamics, $\textit{i.e.}$, the intervalley scattering from K to $\mathrm{\Sigma}$ valley, are representatively shown in (c) $t=0$ fs and (d) $t=100$ fs. (e) Four-dimensional (4D) band structure mapping, $I(E, k_x, k_y, t)$, with the MM showing the band dispersion within the whole Brillouin zone from its center $\Gamma$ to the K valleys at its corners. The same evolution of the excited state is shown for (f) $t=0$ fs and (g) $t=100$ fs, respectively. All the excited states signal are scaled for clarity.}
\label{fig1}
\end{center}
\end{figure}

Fig.~\ref{fig1}(a) depicts the experimental scheme of trARPES employing femtosecond near-infrared (NIR) pump and extreme ultraviolet (XUV) probe pulses 
combined with two types of photoelectron analyzers: a hemispherical analyzer (HA) and a time-of-flight momentum microscope (MM). 
The whole setup allows us to measure the 3D time-dependent electronic structure in a given energy-momentum-plane with high counting statistics using the HA, and alternatively resolve both in-plane momentum directions yielding a 4D photoemission signal $I(E_{kin},k_x,k_y,t)$ of the entire valence band with the MM\cite{puppin2019time, maklar2020quantitative}. Fig.\ref{fig1} (b-d) and (e-g) show snapshots of the 3D and 4D data with 1.55 eV excitation, respectively, at three selected time delays: i) prior to optical excitation, showing the ground-state band structure of WSe$_2$ from the Brillouin zone (BZ) center $\mathrm{\Gamma}$ (only shown in the MM data) to the BZ boundary K points (b,e); ii) upon optical excitation resonant with the A exciton absorption (the first excitonic state), featuring excited-state signal at the $\mathrm{K}$ and $\mathrm{\Sigma}$ valleys (c,f); and iii) at $t=100$ fs after optical excitation, with excited-state signal mostly at the $\mathrm{\Sigma}$ valleys (d,g). 
In the following, we identify the excitonic features in the excited-state photoemission signal and quantify the exciton properties retrieved from the energy, time and momentum dimensions of the 4D trARPES signal.

\paragraph*{Photoemission signature of excitons.}

\begin{figure}
\begin{center}
\includegraphics[width=12cm,keepaspectratio=true]{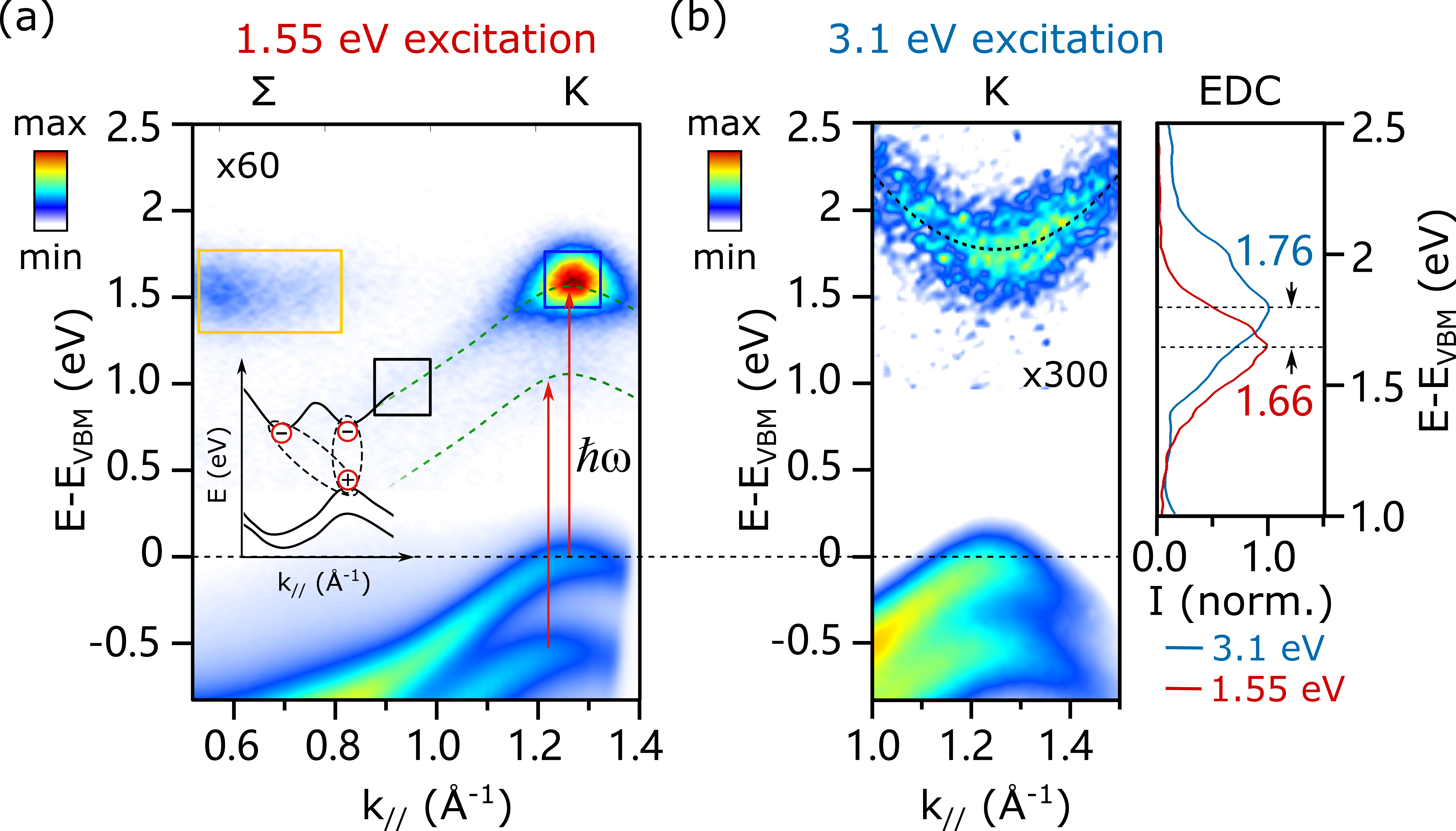}
\caption{\textbf{Signatures of excitons versus quasi-free carriers in trARPES.} (a) Upon the arrival of the pump photons at $\hbar\omega=1.55~ \mathrm{eV}$, the excited states at $\mathrm{K}$ and $\mathrm{\Sigma}$ are populated. During the pump-probe overlap, sideband replica of the two topmost valence bands, VB1 and VB2, are visible (highlighted by the green dashed lines). Inset: schematic of the bright exciton at the K valley and dark exciton at the $\Sigma$ valley. (b) Using the above-band-gap pump at 3.1~eV, the parabolic dispersion of the conduction band at the K valley is clearly observed. As shown in the EDCs at K (right panel), the single-particle (3.1~eV pump photon energy; blue) is $\Delta E\approx 100~\mathrm{meV}$ higher than the excitonic signal at 1.66~eV (1.55 eV pump photon energy; red). The excited state signals are scaled for clarity.}
\label{fig2}
\end{center}
\end{figure}

During the photoemission of an electron bound in an exciton, the electron-hole interaction diminishes, i.e.~the exciton breaks up, as a single-particle photoelectron is detected while a single-particle hole is left behind in the material\cite{weinelt2004dynamics}. To identify the signature of the excitonic electron-hole interaction in photoemission spectroscopy, we compare the signal of excitons with that of single-particle excited states. 
For generating excitons, we excite with 1.55~eV photons ($1/e^2$ bandwidth = 43 meV), in resonance with the low-energy side of the A-excitonic absorption of bulk WSe$_2$\cite{li2014measurement}. 
Fig.\ref{fig2}(a) shows the excited-state signal integrated in the first 25~fs after pump-probe overlap. The data reveals a vertical transition at the K point through an excited-state signal localized in energy and momentum. 
In contrast, the above-band-gap excitation with 3.1~eV photon energy generates a population higher in the conduction band, which rapidly redistributes to all bands and valleys of the lower conduction band (the equivalent scenario applies to the holes in the valence band). Fig.\ref{fig2}(b) shows this excited-state signal in the K valley 100~fs after excitation, where carriers have redistributed in energy and momentum. 
This signal resembles the dispersion of a single-particle band with an effective mass of $m^*=0.55~m_e$, in good agreement with electronic band structure calculations\cite{wickramaratne2014electronic}.
The energetic positions of the excitonic and single-particle states at the K point are determined as the center of mass of the energy distribution curves (EDCs), see Fig.\ref{fig2}(b). The excited-state signal upon resonant excitation of the A exciton is centered $100\pm3~\mathrm{meV}$ below the center of the single-particle band. 
Such a signal below the single-particle band has been predicted as photoemission signature of excitons and the energy difference can be associated with the exciton binding energy $E_b$\cite{rustagi2018photoemission,christiansen2019theory,perfetto2016first,steinhoff2017exciton}. 
A calculation of the A-exciton binding energy in bilayer WSe$_2$ based on the screened Keldysh-like potential (see SI for details) yields $E_b=91.3~\mathrm{meV}$, in very good agreement with the experimental value. 
It is important to note that we retrieve the exciton binding energy directly from measuring the absolute energies of many-body and single-particle states with a single photoemission experiment, in contrast to combining different experimental methods\cite{park2018direct} or by comparing photoemission signals with electronic structure calculations\cite{madeo2020directly}.
The observation that the excitonic binding energy is measurable as energy loss of the photoelectron confirms that the hole final states indeed are identical to single-particle holes, which further implies that the localized hole of the exciton transforms to Bloch-like single-particle states during the photoemission process.  

To set the stage for discussing the exciton dynamics, we emphasize a signal appearing as replicas of the upper (VB1) and lower (VB2) spin-orbit split valence bands in Fig.\ref{fig2}(a) shifted by the photon energy $\hbar\omega_{pump}$. 
This signal only appears during temporal pump-probe overlap and we attribute it to a photon-dressed electronic state due to coherent coupling to the optical driving field.
Since the employed s-polarized pump light (polarization parallel to the sample surface) suppresses laser-assisted photoemission, the experimental configuration selectively probes the coherent excitonic polarization induced by the pump field \cite{perfetto2016first,christiansen2019theory,perfetto2020ultrafast}.  

\paragraph{Formation and decay dynamics of bright excitons.}

\begin{figure}
\begin{center}
\includegraphics[width=16cm,keepaspectratio=true]{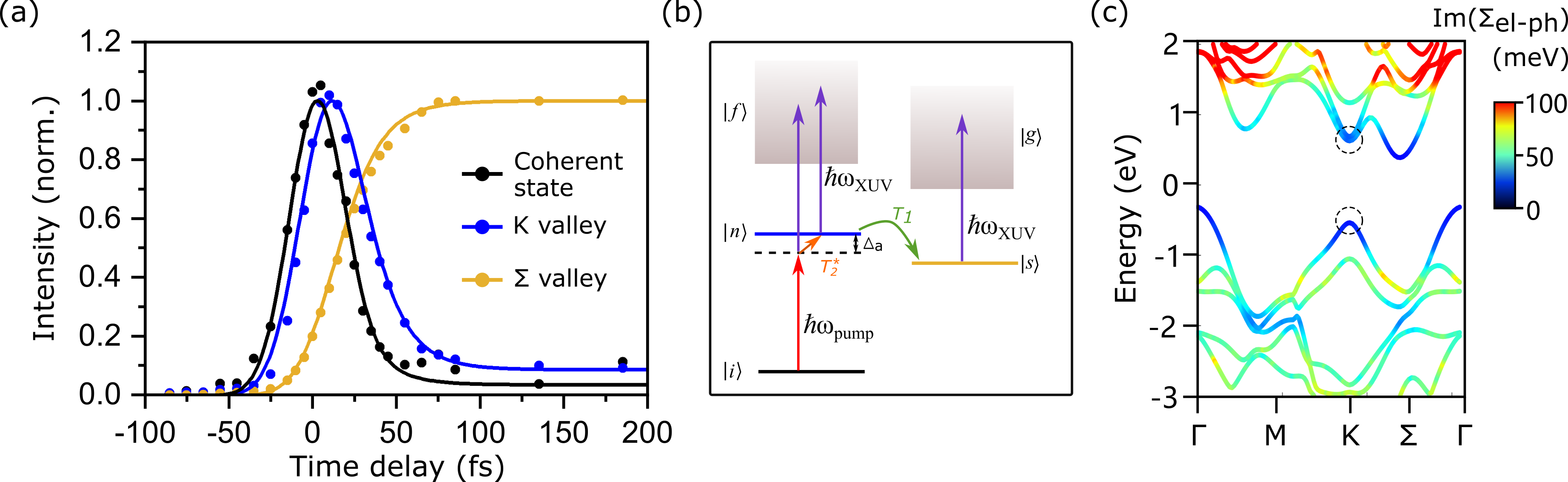}
\caption{\textbf{Exciton dephasing and population dynamics with OBE model fitting.} (a) Normalized photoemission intensity of photon-dressed coherent state (black), bright exciton population at $\mathrm{K}$ (blue) and dark exciton at $\mathrm{\Sigma}$ (yellow) extracted from the three labeled ROIs in Fig.2(a), respectively. The observed time traces are fitted globally with the solution of OBEs (lines). (b) Schematic of the five-level OBE model: the initial state $\ket{i}$ is coherently coupled to the final state $\ket{f}$ through a virtual intermediate state (dashed line), which dephase to the bright excitons at state $\ket{n}$ with the time scale of $T_2^*$; the population at state $\ket{n}$ scatters to the dark state $\ket{s}$ with the time scale of $T_1$; all the excited states are photoionized to the final states $\ket{f}$ and $\ket{g}$, respectively. (c) Imaginary part of the electron-phonon self-energy calculated using density functional perturbation theory. The values at the band extrema at K (circles) are compared with those estimated from the experimental exciton lifetime.}
\label{fig3}
\end{center}
\end{figure}

The bright A excitons at the K point are not the lowest-energy excitons in WSe$_2$ but can relax their energy further by scattering in momentum space. We extract the quasiparticle dynamics within three regions of interest (ROIs) from the trARPES data in Fig.\ref{fig2}(a), representing the coherent excitonic replica of VB1, the excitonic state at the K valley, and the $\Sigma$ valley population.
The respective time traces in Fig.\ref{fig3}(a) reflect three types of quasi-particle dynamics: the dephasing of the coherent excitonic polarization (black), the buildup and relaxation of a bright exciton population at $\mathrm{K}$ (blue) and the carrier accumulation of dark states at $\mathrm{\Sigma}$ (yellow). 

The observed carrier dynamics imply the following microscopic processes as sketched in Fig.\ref{fig3}(b). 
First, the interaction of the initial valence band state $\ket{i}$ with the near-resonant optical light field creates a coherent excitonic polarization (dashed line), which quickly dephases into an optically bright exciton population $\ket{n}$, offset by the pump detuning $\Delta_a$. The decoherence process occurs with the pure dephasing time $T_2^*$. These bright excitons undergo rapid scattering into the optically dark $\mathrm{\Sigma}$-point state $\ket{s}$ on the timescale $T_1$. We model these processes and the photoemission signals from these states into the continuum final states $\ket{f}$ and $\ket{g}$ using a five-level extension to the optical Bloch equations \cite{knoesel1998ultrafast,ueba2007theory} (OBE, see SI).
Based on a multivariate least-squares fitting procedure, we can describe the dynamics of coherent and incoherent exciton contributions, obtaining a coherent exciton dephasing time of $T_2^*=17\pm9$ fs and a population lifetime for the bright A-exciton population of $T_1=18\pm4$ fs. 
The extracted dephasing time corresponds well to  microscopic calculations\cite{raja2018enhancement, selig2016excitonic}. 

To evaluate the mechanism governing the bright exciton scattering, we performed \textit{ab initio} calculations of the single-particle self-energy of WSe$_2$.
At low excitation densities, the electron self-energy is dominated by electron-phonon interaction which is computed using density functional perturbation theory (DFPT), taking into account the electronic screening of the lattice motion (see SI). 
The imaginary part of the momentum-resolved self-energy is shown in Fig.\ref{fig3}(c) encoded by the color scale. 
From the calculation, we obtain $\mathrm{Im(\Sigma_{el-ph})}=13.1~\mathrm{meV}$ at the conduction band minimum and $\mathrm{Im(\Sigma_{h-ph})}=2.6~\mathrm{meV}$ at the valence band maximum of the $\mathrm{K}$ valleys. 
While a rigorous description of exciton-phonon coupling requires treatment on the basis of excitonic eigenstates, in the weak coupling limit, \textit{i.e.}, small self-energy renormalization due to the electron-hole interaction, the exciton-phonon self-energy is dominated by its \textit{incoherent} contribution\cite{marini2008ab}. In this case, the exciton-phonon interaction can be approximated as sum of the single-particle-phonon interactions. Our calculated value $\mathrm{Im(\Sigma_{el-ph})}$+$\mathrm{Im(\Sigma_{h-ph})} \approx16$~meV agrees with the experimental exciton self-energy $\mathrm{Im(\Sigma_{ex})}=18\pm4.8~\mathrm{meV}$ determined according to $\mathrm{Im(\Sigma_{ex})}=\hbar/(2~T_1)$. This agreement with theory shows that the exciton lifetime provides a quantitative measure of the strength of its interaction with the lattice and supports the assumption of a dominating incoherent self-energy contribution.

\paragraph*{Momentum- and real-space distribution of A excitons.}

\begin{figure}
\begin{center}
\includegraphics[width=12cm,keepaspectratio=true]{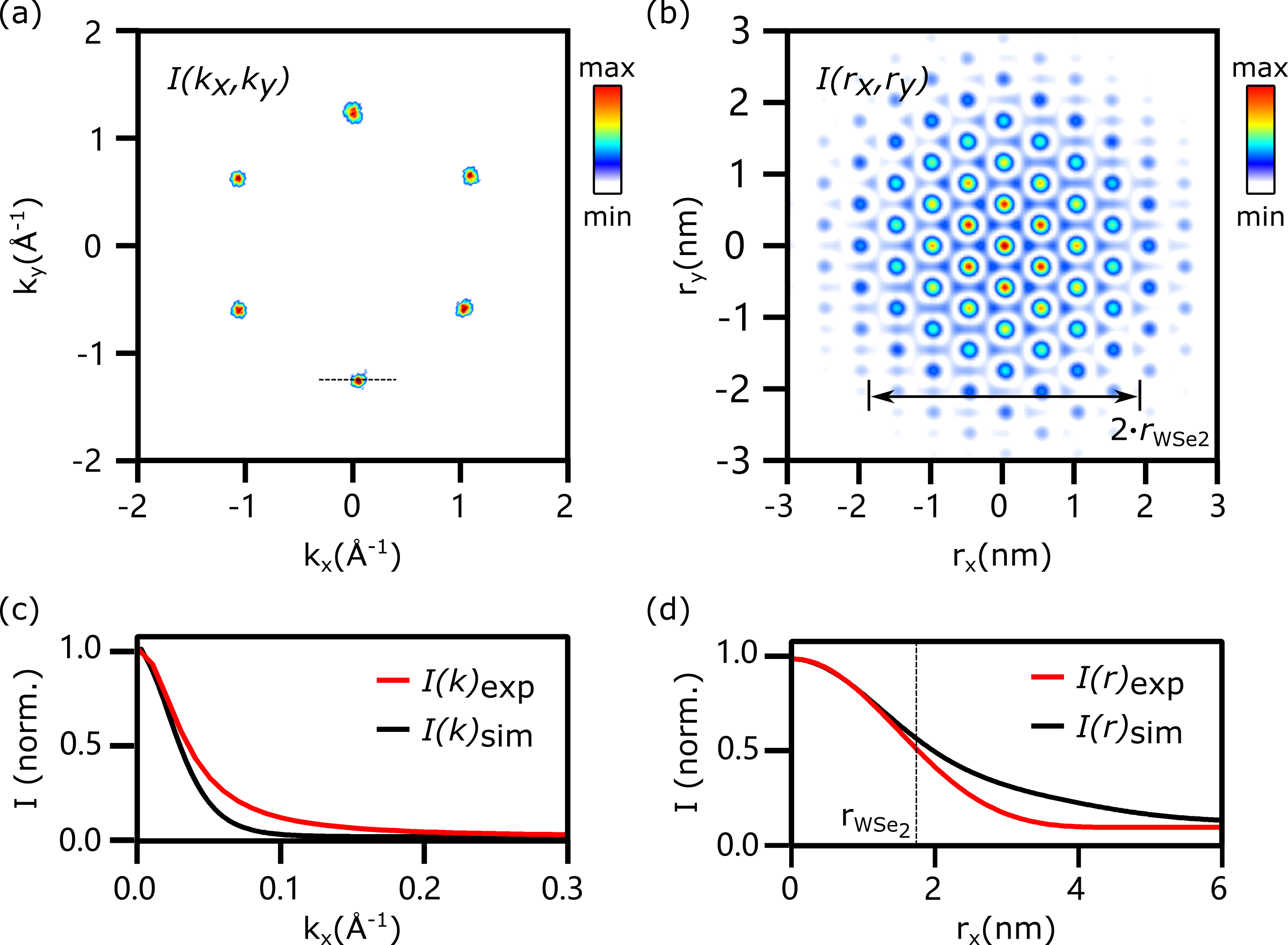}
\caption{\textbf{Momentum- and real-space distribution of A excitons in WSe$_2$.} (a) The early-time momentum distribution of the exciton signal in the six K valleys, $I(k_x,k_y,t=0~\mathrm{fs})$, obtained by energy integration over the CB. (b) 2D Fourier transform of the momentum-resolved photoemission intensity $I(k_x,k_y)$ recovers the real-space image $I(r_x,r_y)$, featuring the electron density distribution of the excitonic wave function. The high frequency oscillations reflect the hexagonal lattice structure. The width of the exciton distribution is indicated by $2\cdot r_\mathrm{WSe_2}$. (c) The momentum distribution curve (MDC) of the bottom K valley (red) extracted along the dashed line in (a), compared with the calculated MDC of the A exciton (black). (d) Experimental and theoretical radial real-space exciton distribution. The exciton Bohr radius $r_\mathrm{WSe_2}$ is indicated with a dashed line. To retrieve the spatial distribution of the exciton, the oscillatory pattern in (b) is removed by Fourier-transforming only one of the K valleys (see SI for details).}
\label{fig4}
\end{center}
\end{figure}

Our 4D trARPES data not only provides the energy-momentum dynamics of excitons but also contains direct amplitude information about exciton wave functions. In Fig.\ref{fig4}(a), we display the early-time excited-state momentum distribution $I(k_x,k_y,t=0~\mathrm{fs})$ of the K valleys, by integrating in energy over the CB. Signals from other valleys are filtered out in order to focus on the A excitons (see SI). 
The total photoemission intensity is proportional to the squared transition dipole matrix element, $|M_{f,i}^\textbf{k}|^2=|\langle\psi_f|\textbf{A}\cdot \textbf{p}| \psi_i\rangle |^2$, which connects the initial state wave function $\psi_i$ to the photoemission final state $\psi_f$, via the polarization operator $\textbf{A}\cdot \textbf{p}$. 
Here, $\textbf{A}$ is the vector potential of the light field and $\textbf{p}$ is the momentum operator. Within the plane wave approximation (PWA) for the final state, the matrix element takes the form 
\begin{equation}
|M_{f,i}^\textbf{k}|^2 \propto |\mathbf{A} \cdot \textbf{k}|^2|\langle e^{i \textbf{k}\cdot \textbf{r}} |\psi_i\rangle |^2
    \label{eq4}
\end{equation}
where $\textbf{k}$ is the wave vector of the photoionized electron. According to Eq.\ref{eq4}, the matrix element is proportional to the amplitude of the Fourier transform (FT) of the initial state wave function. 
Therefore, the momentum distribution of the photoemission signal $I(k_x,k_y)$ can be used to retrieve the real-space probability density of the electron contribution to the two-particle excitonic wave function, i.e., the modulus-squared wave function, $I(r_x,r_y)$, with a suitable assumption for the missing phase information.

A similar reconstruction of electronic wave functions from ARPES spectra has been previously demonstrated for occupied molecular orbitals in the ground state of crystalline organic films and chemisorbed molecular monolayers\cite{offenbacher2015orbital,puschnig2009reconstruction}. Here, we extend this technique into the time domain and apply it to reconstruct the excitonic wave function in WSe$_2$.
Assuming a constant phase profile across the BZ as a lower-limit wave function extension (see SI), we retrieve the exciton probability density via 2D FT as shown in Fig.\ref{fig4}(a-b). 
The reconstruction exhibits a broad isotropic real-space exciton distribution carrying high-frequency oscillations, corresponding to the hexagonal periodic lattice structure of WSe$_2$. 
To resolve the isotropic exciton wave function envelope more clearly, the 1D real-space carrier distribution without the oscillatory pattern is shown in  Fig.\ref{fig4}(d), obtained by FT of only one of the six K valleys, yielding a value of $r_\mathrm{WSe_2}^\mathrm{exp}=1.74\pm0.2$ nm for the excitonic Bohr radius.

To verify the method of reconstructing excitonic wave functions, we performed microscopic calculations of trARPES spectra. The momentum-resolved description of the exciton is based on a many-particle treatment of the Coulomb interaction between electron-hole pairs and the exciton-phonon scattering dynamics\cite{christiansen2019theory} (see SI). 
The momentum distributions of the bright K-excitons calculated within the PWA for the final state is shown in Fig.\ref{fig4}(c). We find a very good agreement to the experimental momentum distribution curve (MDC) taken along the dashed line in Fig.\ref{fig4}(a)), supporting our assumption that the trARPES spectrum contains the fingerprints of the excitonic wave function and justifying the use of the PWA.
Furthermore, the calculated real-space exciton distribution in Fig.~4(d) shows good agreement to our experimental results, yielding a very similar excitonic Bohr radius of $r_\mathrm{WSe_2}^\mathrm{theo}=1.78$ nm. This agreement demonstrates the consistency of the experimentally retrieved exciton binding energy and Bohr radius and additionally suggests the validity of the assumption of a constant phase, which provides the FT-limited (lower-bound) exciton distribution. 
While the excitonic Bloch state is invariant under global and valley-dependent phase renormalization, we find that valley-local phase variations in momentum space can lead to broadening of the exciton probability distribution. In the SI, we reconstruct the real-space exciton density distribution with non-constant intervalley and intravalley phase profiles, where we find a broadened exciton distribution in the case of an intravalley varying phase. Therefore, we note here that the real-space reconstruction of the exciton density with a constant phase is suitable for topologically trivial solid-state wave functions. However, the winding of the phase in topologically non-trivial materials leads to an additional expansion of the carrier density distribution, requiring explicit momentum-dependent phase information. 
In general, the phase of the excitonic wave function might additionally be reconstructed 
through iterative phase retrieval algorithms\cite{jansen2020efficient}.
We envision that future developments will allow retrieving the phase as well as orbital information of excitonic wave functions by utilizing dichroic observables\cite{wiessner2014complete,beaulieu2020revealing,schuler2021bloch} in trARPES.

In this work, we provide a comprehensive experimental characterization of an excitonic state with trARPES. 
The interactions governing the formation of this prototypical many-body state are observable as energy renormalization in comparison to single-particle states, while its interaction strength with other quasi-particles is reflected in the excited state's lifetime. 
These quantities are intimately connected to the real and imaginary parts of the many-body state's self-energy and our approach establishes experimental access to these elusive quantities. 
Moreover, we retrieve real-space information of the excitons by Fourier transform of its momentum distribution, establishing the measurement of wave function properties of transient many-body states with 4D photoemission spectroscopy.
Our approach is applicable to all exciton species occurring in a wide range of inorganic and organic semiconductors, van der Waals heterostructures and devices. Its extension to other many-body quasi-particles in solids appears straightforward.

\noindent
\textbf{Data availability}
We provide the full experimental dataset as well as the details of the data analysis on the data repository Zenodo. Also, we provide the source code of our data analytics on GitHub.

\bibliographystyle{cite}

\textbf{Acknowledgments:} This work was funded by the Max Planck Society,the European Research Council (ERC) under the European Union’s Horizon 2020 research and innovation and the H2020-EU.1.2.1.~FET Open programs (Grant Nos. ERC-2015-CoG-682843, ERC-2015-AdG-694097, and OPTOlogic 899794), the Max Planck Society’s Research Network BiGmax on Big-Data-Driven Materials-Science, and the German Research Foundation (DFG) within the Emmy Noether program (Grant No. RE 3977/1), through SFB 951 "Hybrid Inorganic/Organic Systems for Opto-Electronics (HIOS)" (Project No. 182087777, projects B12 and B17), the SFB/TRR 227 "Ultrafast Spin Dynamics" (projects A09 and B07), the Research Unit FOR~1700 "Atomic Wires" (project E5), and the Priority Program SPP~2244 (project 443366970). 
D.C.~thanks the graduate school Advanced Materials (SFB 951) for support.
S.B.~acknowledges financial support from the NSERC-Banting Postdoctoral Fellowships Program. T.P.~acknowledges financial support from the Alexander von Humboldt Foundation.
\textbf{Author contributions:} S.D., M.P., S.B., T.P., C.W.N., M.D., Y.D., Y.W.W., M.W., L.R., and R.E.~designed, prepared, and performed the experiment. T.P., M.P, R.P.X., and R.E.~performed the OBE fitting. D.C., M.S., E.M., and A.K.~performed the calculation of trARPES spectrum. H.H.~and A.R.~calculated the single-particle self-energy. R.P.X.~developed the 4D data processing code. All authors contributed to the manuscript. 
\textbf{Competing Interests:} The authors declare that they have no conflict of interest.

\clearpage
\end{document}


\renewcommand\thefigure{S.\arabic{figure}}
\renewcommand\thetable{S.\arabic{table}}

\baselineskip24pt


\maketitle


\begin{sciabstract}
\end{sciabstract}


\noindent
\begin{LARGE}  
Supplementary Notes  
\end{LARGE} 

\noindent
This section includes the experimental methods, procedure of reconstructing the real-space excitonic wavefunction and energy calibration of trARPES spectrum.

\noindent
\ul{\textbf{Experimental Methods}}

\noindent
\textbf{Time- and angle-resolved photoemission spectroscopy.}
The whole setup and the computational workflow for data processing have been described in detail elsewhere \cite{puppin2019time, maklar2020quantitative,xian2019open}. Our laser system is a home-built optical parametric chirped-pulse amplifier (OPCPA) delivering 15 W ($\lambda_{pump}=800~\mathrm{nm}$) at 500 kHz repetition rate \cite{puppin2015500}. The major part (80\%) of the OPCPA output is used to drive high-order harmonic generation (HHG) by tightly focusing the second harmonic of the laser pulses (400 nm) onto a dense Argon gas jet. 
Out of the generated XUV frequency comb, a single harmonic (7th order, 21.7 eV) is isolated by a combination of a mulilayer mirror and propagation through a 400 nm thick Sn metallic filter. 
The remaining part of the OPCPA output serves as the optical pump beam, with a transform-limited pulse duration of 35 fs. 
Another pump wavelength used in the experiment, $\lambda_{pump}=400~\mathrm{nm}$, is the frequency-doubled fundamental pump light generated using a barium borate crystal.
In the measurement, the pump fluence of 800 nm is $F_{800}=1.3 ~\mathrm{m J/cm^2}$ and that of 400 nm is $F_{400}=85 ~\mathrm{\mu J/cm^2}$. 
All measurements are performed at room temperature. 
The optical pump and probe beams are focused at the sample position in the ultra-high vacuum chamber which is equipped with two type of photoelectron analyzers, a conventional hemispherical analyser (HA) from SPECS GmbH and a novel time-of-flight momentum microscope (MM) from Surface Concept and SPECS GmbH. 
This combination of complimentary electron analyzers enables high quality and efficient data collection within the full Brillouin zone (with the MM) and regions of interests (with the HA), and the exploration of ultrafast dynamics.
The MM collects photoelectrons within a wide emission angle using an extraction lens, simultaneously recording the photoemission spectrum at both in-plane momentum directions $k_x$ and $k_y$ using a delay line detector. 
In a pump-probe scheme, the MM thus directly provides the 4D-dimensional photoemission intensity I($E_{kin},k_x,k_y,t$) in an efficient way as shown in the Fig.1. 
In contrast, the HA yields a single energy-momentum cut in a fixed experimental configuration, effectively allowing for a much higher electron detection rate within a particular momentum range, which allows us to investigate the delicate quasiparticle scattering dynamics with enhanced signal-to-noise ratio. 
In our experiment, the energy axis of trARPES spectra are aligned with the ground state of valence band maximum (VBM) at the K valley. 

\noindent
\textbf{Sample preparation.}
Bulk WSe$_2$ is a purchased crystal from HQ Graphene, which is firstly glued on top of a copper sample holder and then cleaved at room temperature and a base pressure of 5x10$^{-11}$ mbar. The sample is further handled by a 6-axis manipulator (SPECS GmbH) for trARPES measurements.

\noindent
\ul{\textbf{Determination of the excitonic distribution function}}

\noindent
\textbf{Matrix element effects in the ARPES spectrum}
\noindent
Within the sudden approximation, the intensity of the ARPES spectrum can be written as the product of a transition matrix element $M_{f,i}^\textbf{k}$, the one-electron removal spectral function $A(k,\omega)$ and the electronic distribution function $f(k,\omega)$: 
\begin{equation}
   I(k,\omega)\propto\frac{2\pi}{\hbar}|M_{f,i}^\textbf{k}|^2A(k,\omega)f(k,\omega)
\end{equation}
As discussed in the main text, the matrix element is defined as a transition between the initial and final state wave functions$ |M_{f,i}^\textbf{k}|^2=|\langle\psi_f|\textbf{A}\cdot \textbf{p}| \psi_i\rangle |^2$, mediated by the vector potential of the exciting electromagnetic field $\textbf{A}$ and the momentum operator \textbf{p}. Assuming the final state as a plane wave, this component can be simplified as $|M_{f,i}^\textbf{k}|^2 \propto |\mathbf{A} \cdot \textbf{k}|^2|\langle e^{i\textbf{k}\cdot \textbf{r}} |\psi_i\rangle |^2$, containing two elements: (1) the polarization factor $|\textbf{A}\cdot \textbf{k}|$ between the potential of the optical field $\textbf{A}$ and the photoelectron wave vector $\textbf{k}$, and (2) the Fourier transformation of the initial state wave function $\psi_i(k)$. The first term is largely determined by the experimental geometry, leading to a slight modulation of the spectral weight based on the projection of polarization to the final state wave vector\cite{moser2017experimentalist}. Therefore, it would unavoidable introduce a intensity modulation in the practical use of hemispherical electron analyser, because of the sample rotations during the momentum scan. In our setup, the momentum microscope eliminates the experiment-induced intensity variation by collecting the photo-ionized electrons of a wide angle range, such that the momentum map can be achieved in a fixed experimental geometry. The matrix element component is naturally encoded with the information of initial state as discussed in the main text. The photoemission intensity modulation due to the orbital texture has been demonstrated in the recent study of time-reversal dichroism in ARPES\cite{beaulieu2020revealing}. Under the plane wave assumption (PWA) of final states, the real-space carrier reconstruction using its momentum distribution achieves the experimentally long-pursued goal: mapping the fundamental shape and size of electronic wavefunction. Nevertheless, the validity of the Fourier imaging approach, more specifically speaking, the PWA of the final states has been continuously debated, which is considered as applicable under the favorable conditions of simple orbital components and high photon energy \cite{bradshaw2015molecular}. The good agreement of the microscopy calculation of trARPES spectrum using PWA of final states with the experimental results, shown in the main text Fig.4, supports the viability of real-space reconstruction of exciton in our system.                

\paragraph{The data preparation procedure}

\begin{figure}
\begin{center}
\includegraphics[width=15.8cm,keepaspectratio=true]{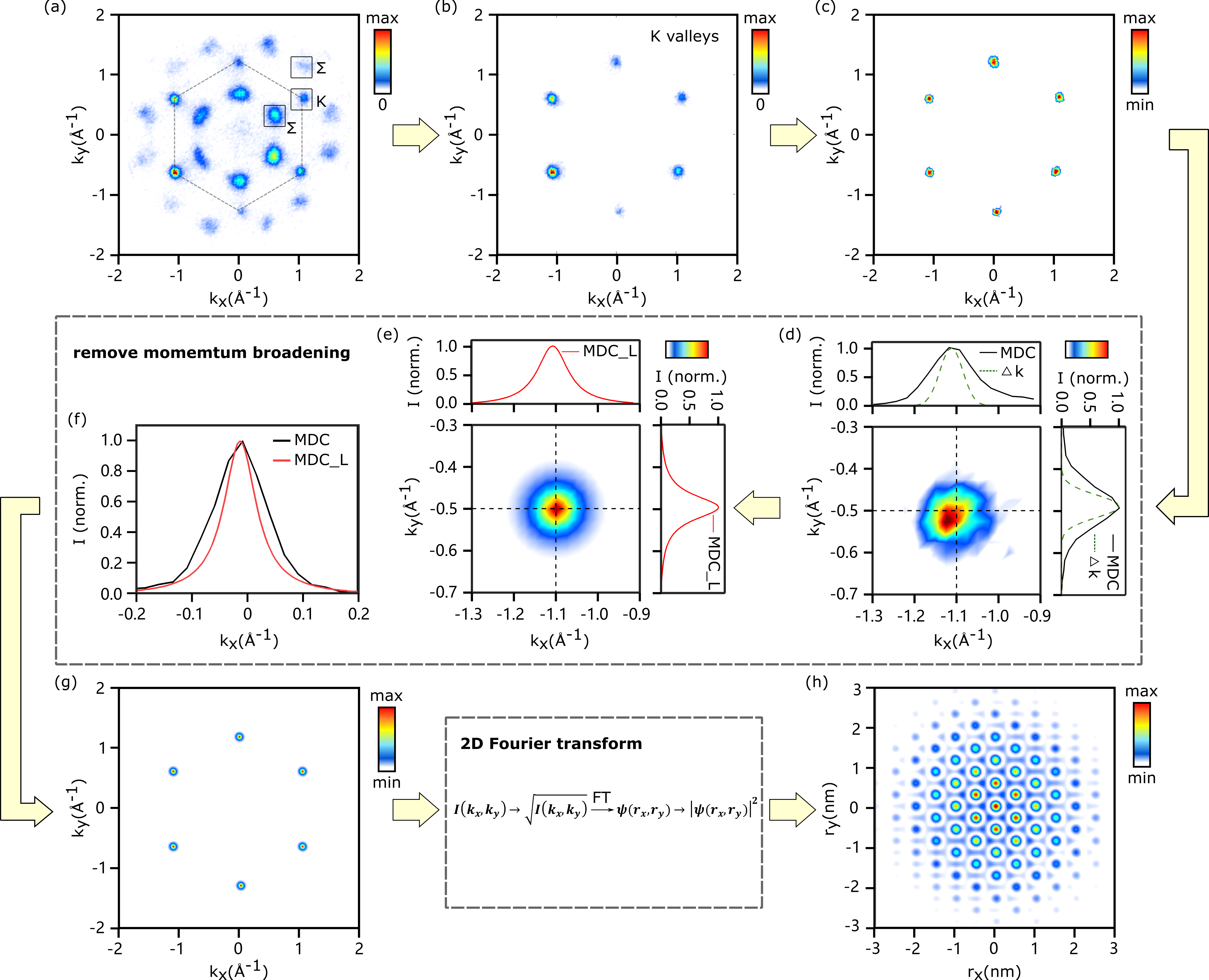}
\caption{\textbf{Momentum and real-space carrier distribution map of WSe$_2$}. (a) Start from the momentum map of excited state electrons which includes six K valleys in the first BZ (dash line labelled) and $\Sigma$ valleys in the first and second BZ. (b) The six K valleys are selectively shown. (c) The intensity at each K valley are normalized to remove the polarization factor. (d) Zooming to one of the K valley, the MDCs ($k_x$ and $k_y$ cuts at top and right panel) are selected at the center of carrier distribution. The superimposed momentum response function (green dashed Gaussian shape with FHWM=$0.063~\mathrm{\AA}^{-1}$) characterises the momentum resolution of the setup. (e) After deconvolving the momentum broadening factor, the intrinsic carrier distribution is presented as 2D Lorentzian shape. (f) The comparison of raw MDC data and extracted Lorentzian MDC. (g) The momentum map is then 2D Fourier transferred to real-space distribution (h).}
\label{figS2}
\end{center}
\end{figure}

The momentum maps as shown in Fig.4 (a) in the main text are prepared and consequently Fourier transformed to real space following three step: (i) isolation of the K valleys from the trARPES data (time delay is selected at time zero) by applying a momentum mask, (ii) removal of the momentum broadening effect, mainly from the momentum resolution of setup, employing a deconvolution scheme, and (iii) 2D Fourier transformation (FT) of the momentum map to real space. 
Fig.~\ref{figS2} shows the detailed data processing procedures step by step. By integrating the energy axis within the conduction band region, we obtain the conduction band momentum map showing the excited state carrier distribution localized at the conduction band minima around the K and $\Sigma$ valleys in the first Brillouin zone (BZ), and some of the $\Sigma$ valleys of the second BZ (Fig.~\ref{figS2}(a)). By applying a circular momentum mask, we isolate the six K valleys as shown in Fig.~\ref{figS2}(b). Next, the population intensity at each K valley is normalized by the local maximum to remove the geometric polarization factor (Fig.~\ref{figS2}(c)), which comes from the coupling between light field and photoionized electron determined by the experimental geometry. In this measurement, we used linearly polarized light at $65^\circ$ incidence angle with respect to the sample surface. 
In step (ii), the experimental momentum resolution effect is deconvolved from the measured data to obtain the intrinsic momentum distribution. Our momentum resolution of METIS is, $\delta k = 0.063~\mathrm{\AA}^{-1}$, determined by the grid edge at back focusing plane. The MDC of the raw data (Fig.~\ref{figS2}(d)) was maximum likelihood fitted with the convolved function of a Gaussian (FHWM=$\delta k$) and a Lorentzian function representing the intrinsic spectral function (Fig.~\ref{figS2}(e) and (f)). Finally, in step(iii), the real-space carrier distribution (Fig.~\ref{figS2}(h)) is obtained by applying the 2D FT to the momentum map (Fig.~\ref{figS2}(g)). Noted, we present the intrinsic momentum distribution of six K valleys with the averaged lineshape after removing the momentum resolution effect. Within the 2D FT process, we transfer the square root of the intensity of the 2D momentum map $\sqrt{I(k_x,k_y)}$ to the real-space wavefunction $\psi(r_x,r_y)$ and then square the wave function to present the probability $|\psi(r_x,r_y)|^2$. The neglected phase information will be discussed below.

In analogy to the electron Bohr radius in hydrogen atom\cite{griffiths2018introduction}, we estimate the exciton Bohr radius based on the average radius of exciton distribution $\langle r^2 \rangle=|r\cdot\psi(r)|^2$. The 1D real-space exciton distribution $|\psi(r)|^2$ is shown in the main context Fig.4(d) and the exciton Bohr radius $r_\mathrm{WSe_2}$ is defined as the distance of the peak intensity in $\langle r^2 \rangle$. By calculating the exciton distribution at six K valleys, we obtain the exciton Bohr radius $r_\mathrm{WSe_2}=1.74\pm0.2$ nm, which the standard deviation is a measure of variability between valleys.

\paragraph{Momentum dependent phase profile of exciton wave function}

In the FT procedure described in the main text, we assumed a constant phase profile of the excitonic wave function, which yields a lower limit for the excitonic distribution function. This relation will be rationalized in this section, showing that while the excitonic Bloch state is invariant under global and valley-dependent phase shift, local phase variations in momentum space only lead to broadening in the spatial exciton probability distribution. 

While more advanced approaches such as phase-retrieval schemes have been employed to reconstruct the amplitude and phase of ARPES spectra in iterative algorithms \cite{kliuiev2018algorithms}, here we discuss the influence of inter- and intravalley-dependent phase variations based on symmetry considerations and simulations. With linearly polarized excitation, bulk WSe$_2$ preserves both time reversal symmetry and spatial inversion symmetry. The time reversal operator $\hat{T}$ introduces a phase reversal between K and K’ point, which suggests the Bloch wave function can be written as $\psi_K (r)=e^{ik\cdot r} \phi_K (r)e^{i\theta}$ and $\psi_{K'} (r)=e^{ik\cdot r} \phi_{K'} (r)e^{-i\theta}$, respectively. 

\begin{figure}
\begin{center}
\includegraphics[width=15.8cm,keepaspectratio=true]{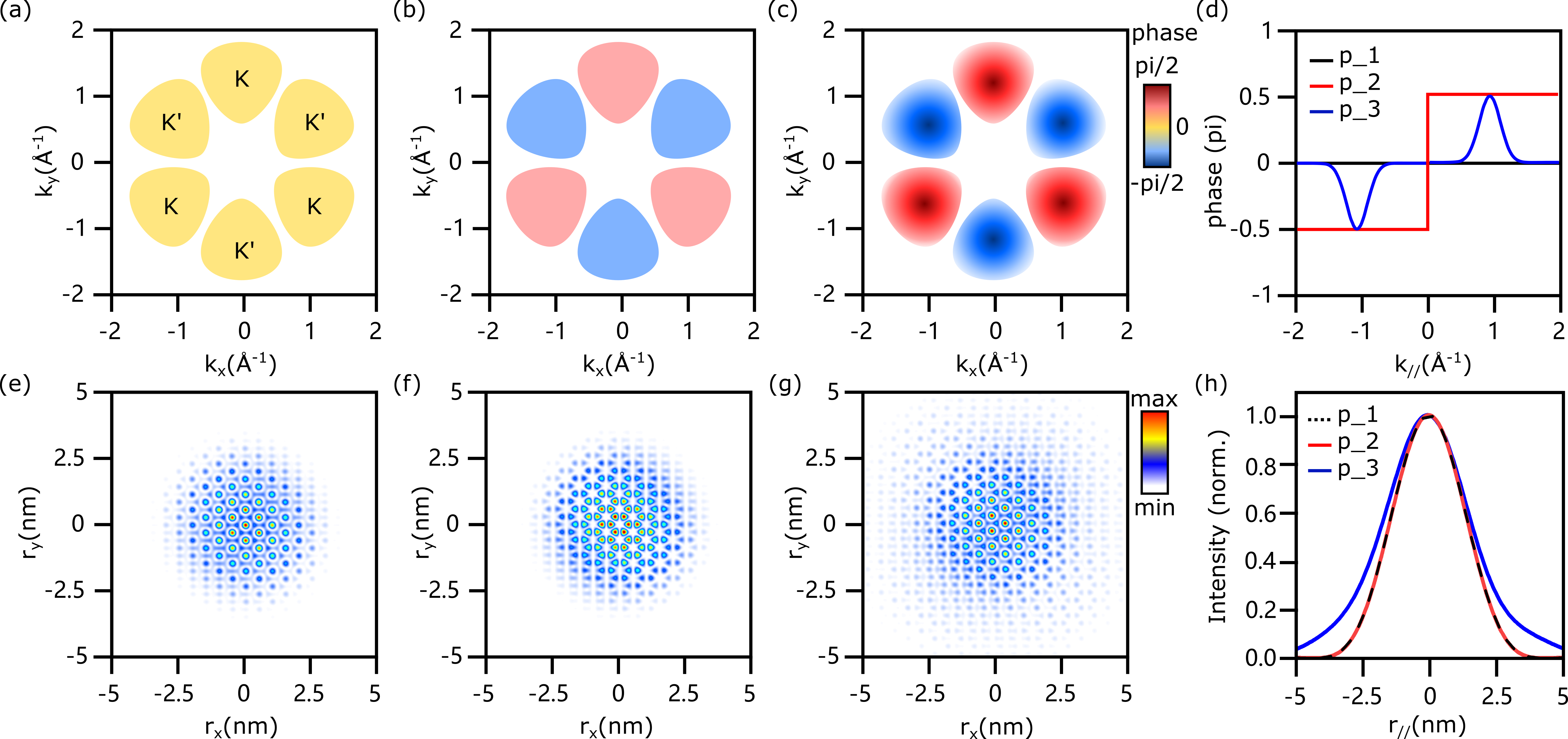}
\caption{\textbf{Momentum dependent phase profile of A excitons}. Three representative phase profiles are prepared as (a) constant, (b) intervalley alternating and (c) inter- and intravalley varying  in momentum space. The corresponding phase information within these three scenarios are summarized in (d). The real-space exciton density distribution of bulk WSe$_2$ (e), (f) and (g) are obtained via FT of the experimentally measured amplitude profile of the wave function with the phase profiles (a), (b) and (c), respectively. (h) Normalized probability distribution of exciton density for the three different phase scenarios, showing the influence of local phase variations on the envelop function.}
\label{figS3}
\end{center}
\end{figure}

To fulfill these conditions, we construct a periodical varying phase mask of $e^{i\theta}$ and $e^{-i\theta}$ centered at the K and K' valleys, respectively (Fig.~\ref{figS3}(b)), and apply it to the amplitude profile from our measurement (square root of the momentum map) before reconstructing the real-space image. The corresponding exciton density distribution can be found in Fig.~\ref{figS3}(f). Compared with the momentum-independent phase profile (Fig.~\ref{figS3}(a)) which was applied for the evaluation in the main text, we find no influence on the width of the exciton distribution (Fig.~\ref{figS3}(h)). Note that the high frequency oscillations under the envelope function vary with the intervalley phase reversal, which can be regarded as analogous to the interference pattern changing with the relative phase differences between two coherent emitters (K and K' valleys). 

A more realistic phase profile might contain on top of an intervalley alternating phase contribution also a non-constant intravalley phase profile, as exemplary shown in Fig.~\ref{figS3}(c). Importantly, we find that such a non-constant inter-valley phase profile can only lead to a broadening of the exciton distribution, see Fig.~\ref{figS3}(g). While these considerations show the importance of the local phase profile for a detailed assessment of the excitonic wave function, the assumption of a flat phase profile can be used to estimate a lower limit. Note that the phase profiles we prepared are three representative cases of inter- and intra-valley dependent phase information.


\noindent
\ul{\textbf{Energy alignment of trARPES spectrum}}

\begin{figure}
\begin{center}
\includegraphics[width=15.8cm,keepaspectratio=true]{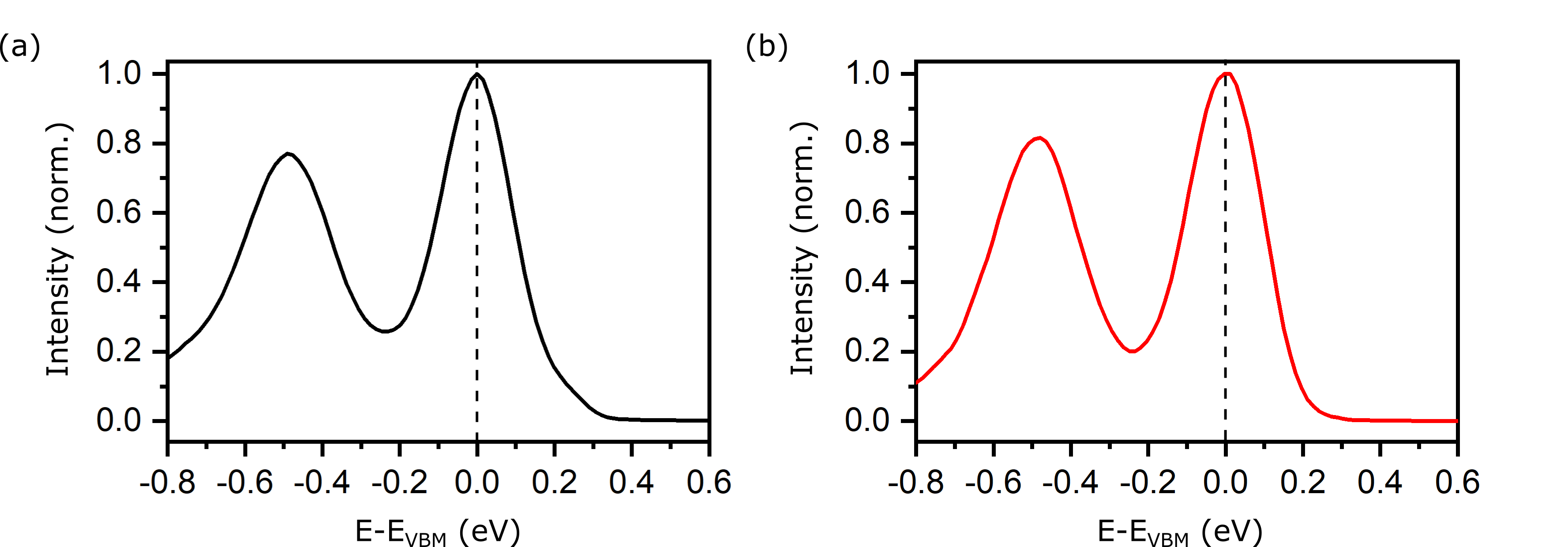}
\caption{\textbf{Energy calibration of trARPES spectrum}. (a),(b) EDC of the K valley at $t=-1~\mathrm{ps}$ before 800 nm and 400 nm excitation, respectively}
\label{figS4}
\end{center}
\end{figure}

\noindent
 In our experiment, the energy axis of trARPES spectra are aligned with the ground state of valence band maximum (VBM) at the K valley. Fig.~\ref{figS4} shows the energy distribution curve (EDC) at $t=-1~\mathrm{ps}$ before 800 nm (black) and 400 nm pump (red), respectively.

\newpage
\noindent
\begin{LARGE}  
Supplementary Methods
\end{LARGE} 

\noindent
This section includes the optical Bloch equation model, single-particle self-energy calculation and microscopic calculation of exciton in trARPES.

\noindent
\ul{\textbf{Optical Bloch equation fitting}}

\noindent
Optical Bloch equation (OBE) modelling has been employed to describe two-photon photoemission from metallic surface states \cite{weinelt2002time,hertel1996ultrafast}. In general, the use of OBE in solid state systems and in particular in photoemission hinges on the important result that the coherent excitation of a quasi-continuum of states may be treated in the same way as the incoherent limit of a pure two level system, significantly simplifying the full harmonic description of the quantum process \cite{hertel2015atoms}.
It is therefore possible to apply OBE to photoemission processes within semiconductors when the conditions allow schematizing the system as a set of atomic levels. In particular, it is well suited for pump excitation wavelengths very close to the direct bandgap of the semiconductor. In this conditions, hot carriers have very little excess energy and intra-valley relaxation dynamics can be neglected.

OBE offer a tool to disentangle the formation and decay of a coherent polarized state of the solid from the intrinsic lifetime of electronic states. This allows for a much more precise assessment of intrinsic lifetimes than models based on rate equations that ignore the decoherence processes, and tend to grossly overestimate the real values, especially when the lifetime is close to, or even shorter than, the temporal resolution, i.e. pump-probe pulse cross-correlation.
Here we explain how the three level OBE model is extended in order to determine the dephasing time of two-photon photoemission and the intrinsic lifetime of bright excitons at the WSe$_2$ K-valley.  

\paragraph{Model construction}

For the derivation of the equations for the 3 level system, we use the formula described by Martin Weinelt's \cite{weinelt2002time}. 
To simplify the formalism, the energy variables are expressed in terms of “detunings”, i.e.:
\[\Delta_a=\hbar \omega_a-(E_n-E_i )\]
\[\Delta_b=\hbar \omega_b-(E_f-E_n )\]
These quantities contain all the relevant energy information for the model and are zero when a transition is resonant to the photon energy of the pump, $\hbar \omega_a$, or the probe, $\hbar \omega_b$: $E_i$ is the energy of the initial state, $E_n$ the one of the intermediate state of interest, and $E_f$ is the final, detected state energy. 
In our case, $\Delta_a$ is zero when the pump energy is resonant with the bright A-exciton populating the K valley, and $\Delta_b$ is zero when a value of energy of the continuum is resonant with the photoemission from the intermediate state due to the probe photon. 
Note that $\Delta_{a,b}\neq0$ can be produced by two different mechanisms: either $\hbar \omega_{a,b}$ is changed while $(E_{n,f}-E_{i,n})$ remains constant (which we may call \textit{wavelength detuning}) or, alternatively, the effective energy gap for vertical transitions $(E_{n,f}-E_{i,n})$ is changed, while $\hbar \omega_{a,b}$ is constant (which we may call \textit{energy gap detuning}).

In the application of OBE to photoemission, $\Delta_a$ and $\Delta_b$ are used with a different conceptual approach. As the pump creates transitions between states in the solid, $\Delta_a\neq0$ is intended to arise either from wavelength detuning, or from energy gap detuning between Bloch states in the solid (for example, a change in the bandgap due to temperature variation). 
$\Delta_b$ refers, instead, to vacuum states whose energy can vary continuously. Because of this, $\Delta_b\neq0$ is used, in the following, as a way to map the measured ARPES spectra in the final detected state: $\hbar \omega_{b}$ and $E_f$ are considered fixed, and $\Delta_b\neq0$ allows to produce transitions from states energetically above or below $E_n$ to the observed final state. The energy-resolved spectrum can therefore be obtained by continuously varying $\Delta_b$.

The derivation of OBE is based on the Liouville-Von Neumann equation:
\begin{equation} \label{eq:Liouville}
\rho_{kl}=-i[\hat{H}_0+\hat{V},\hat{\rho}_{kl}]-\Gamma_{kl} \rho_{kl}
\end{equation}
The equation describes the evolution of the density operator under a perturbation $\hat{V}$ with unperturbed Hamiltonian $\hat{H}_0$. Note that $\hbar$ has been set to 1, as will be in the following.
The perturbing fields are described in the dipole approximation by the quantities:
\begin{equation} \label{eq:pump_envelope}
    p_a(t) = E_a(t) \Braket{i|D|n}
\end{equation}
\begin{equation} \label{eq:probe_envelope}
    p_b(t) = E_b(t) \Braket{n|D|f}
\end{equation}
where the temporal envelope functions $E_{a,b}(t)$ of both pump and probe pulses are assumed to be Gaussian distributions and $D$ is the dipole operator. 
In order to account for intrinsic population decay and decoherence term in eq.\ref{eq:Liouville}, we add the damping operator $\hat{\Gamma}$, that can be represented in matrix form as: 
\begin{equation}\label{eq:GammaMat}
    \hat{\Gamma}=\begin{pmatrix}
0 & \dfrac{\Gamma_n}{2}+\Gamma_n^*+\Gamma_i^* & \Gamma_i^*+\Gamma_f^*\\
\dfrac{\Gamma_n}{2}+\Gamma_n^*+\Gamma_i^* & \Gamma_n & \dfrac{\Gamma_n}{2}+\Gamma_n^*+\Gamma_f^*\\
\Gamma_i^*+\Gamma_f^* & \dfrac{\Gamma_n}{2}+\Gamma_n^*+\Gamma_f^* & 0
\end{pmatrix}
\end{equation}
The terms $\Gamma_j$ and $\Gamma_j^*$ describe the decay and the dephasing rates of state $\textit{j}$, respectively.
In eq.\ref{eq:GammaMat}, the intrinsic lifetime of excitons can be characterised by the diagonal term $\hat{\Gamma}_n$, describing the population decay rate. The initial and the final state are assumed to have an infinite lifetime, i.e., $\Gamma_i= \Gamma_f = 0$. On the other hand, the off-diagonal terms of intermediate state contains half of the decay rate and the complex part $\Gamma_n^*+\Gamma_i^*$, which is the so-called “pure-dephasing rate”, describing the decay of quantum coherence between the levels $n$ and $i$.

\begin{figure}
\begin{center}
 \includegraphics[width=15.5cm,keepaspectratio=true]{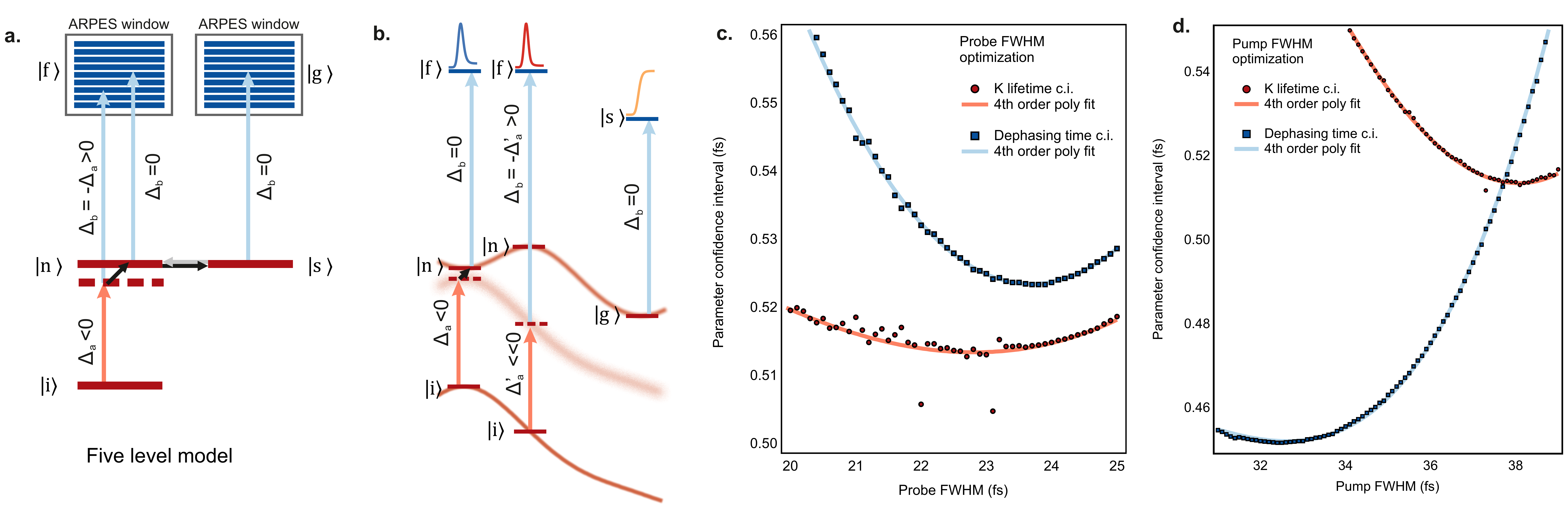}
 \caption{\textbf{Schematics and optimization of fit parameters}. 
 (a) Schematic of the five level model in the presence of pump detuning. The red levels represent the energy levels, the blue ones the continuum states of ARPES final states. The dashed line represents the virtual intermediate state for the coherent two-photon process. The orange arrow represents the pump excitation, the light blue arrow the probe-induced transition. The black diagonal arrow represents the dephasing of the coherence in an excited state population. The black horizontal arrow represents scattering to the two level subsystem ($g$ and $s$), while the grey one represents the small fraction of back-scattered carriers.  (b) Schematic view of the procedure used to improve the fit precision. The  bottom (top) horizontal orange line represents the dispersion of the VB (CB), while the blurry replica represents the transient coherent state. The data for K and $\Sigma$ are extracted in the corresponding momentum positions ($k=1.2\mathrm{\AA}^{-1}$ and $k=0.6\mathrm{\AA}^{-1}$) and fitted with the OBE solved for small negative pump detuning and zero probe detuning (blue and yellow curve). Simultaneously, the data extracted from the Floquet state at $k=0.9\mathrm{\AA}^{-1}$ is fitted with the OBE solution for large negative pump detuning ($\Delta_a'$) and corresponding positive probe detuning $\Delta_b=-\Delta_a'$ (red curve). (c) Variation of the confidence interval relative to the K-exciton lifetime (red circles) and Dephasing time (blue squares) as a function of the probe FWHM. The curves have been fitted with 4th order polynomials, removing the effects of numerical noise allowing the determination of the position of the minima with high precision.  (d) Similar, but as a function of the pump FWHM.
}
 \label{figS1_1}
\end{center}
\end{figure}

By substituting eq.\ref{eq:pump_envelope}-\ref{eq:GammaMat} in eq.\ref{eq:Liouville}, and applying the rotating wave approximation (neglecting high frequency oscillating terms), we obtain the following optical Bloch equations:
\begin{align}
        \dot{\rho}_{ii} &=+\mathfrak{Im} \left( p_a^* \rho_{in}^{(1)} \right) \label{eq:OBE3_1}\\
        \dot{\rho}_{nn} &=-\mathfrak{Im} \left( p_a^* \rho_{in}^{(1)} \right) +\mathfrak{Im} \left( p_b^* \rho_{nf}^{(2)} \right) - \Gamma_n \rho_{nn} \label{eq:OBE3_2}\\
        \dot{\rho}_{ff} &= - \mathfrak{Im} \left( p_b^* \rho_{nf}^{(2)} \right) \label{eq:OBE3_3}\\
        \dot{\rho}_{in}^{(1)} &=-i\Delta_a\rho_{in}^{(1)}-\dfrac{i}{2}p_b^* \rho_{if}^{(3)}+\dfrac{i}{2}p_a(\rho_{nn}-\rho_{ii})-\left( \dfrac{\Gamma_n}{2}+\Gamma_i^*+\Gamma_n^*\right) \rho_{in}^{(1)} \label{eq:OBE3_4}\\
        \dot{\rho}_{nf}^{(2)} &=-i\Delta_b\rho_{nf}^{(2)}+\dfrac{i}{2}p_a^* \rho_{if}^{(3)}+\dfrac{i}{2}p_b(\rho_{ff}-\rho_{nn})-\left( \dfrac{\Gamma_n}{2}+\Gamma_n^*+\Gamma_f^*\right) \rho_{nf}^{(2)} \label{eq:OBE3_5}\\
        \dot{\rho}_{if}^{(3)} &=-i(\Delta_a+\Delta_b)\rho_{if}^{(3)}+\dfrac{i}{2}p_a^* \rho_{nf}^{(2)}+\dfrac{i}{2}p_b\rho_{in}^{(1)}-\left(\Gamma_i^*+\Gamma_f^*\right) \rho_{if}^{(3)} \label{eq:OBE3_6}
\end{align}
where we have set:
\begin{align*}
    \rho_{in}^{(1)} &= e^{-i\omega_a t}\rho_{in}\\
    \rho_{nf}^{(2)} &= e^{-i\omega_b t}\rho_{nf}\\
    \rho_{if}^{(3)} &= e^{-i(\omega_a+\omega_b) t}\rho_{if}
\end{align*}
This is the set of equations needed to describe a three level system. To adapt the formalism to include K-$\mathrm{\Sigma}$ incoherent scattering, we need to create an additional two level system coupled to the main three level system via the K decay rate $\Gamma_n$. We have to build an equation for the $\mathrm{\Sigma}$ state (subscript $s$), one for the photoemitted final state (subscript $g$), and one for the corresponding coherence (subscript $sg$). To meet the experimental reality, we also insert a backscattering parameter $\mathfrak{B}$, allowing to account for all mechanisms transferring population from $\mathrm{\Sigma}$ to K (eq. \ref{eq:OBE3_2} has to be modified, see eq. \ref{eq:OBE5_2}). For a schematic of the 5-level system, see Fig.~\ref{figS1_1}(a).
\begin{align}
        \dot{\rho}_{ii} &=+\mathfrak{Im} \left( p_a^* \rho_{in}^{(1)} \right) \label{eq:OBE5_1}\\
        \dot{\rho}_{nn} &=-\mathfrak{Im} \left( p_a^* \rho_{in}^{(1)} \right) +\mathfrak{Im} \left( p_b^* \rho_{nf}^{(2)} \right) - \Gamma_n \rho_{nn} {\color{red} + \Gamma_n \mathfrak{B} \rho_{ss}} \label{eq:OBE5_2}\\
        \dot{\rho}_{ff} &= - \mathfrak{Im} \left( p_b^* \rho_{nf}^{(2)} \right) \label{eq:OBE5_3}\\
        \dot{\rho}_{in}^{(1)} &=-i\Delta_a\rho_{in}^{(1)}-\dfrac{i}{2}p_b^* \rho_{if}^{(3)}+\dfrac{i}{2}p_a(\rho_{nn}-\rho_{ii})-\left( \dfrac{\Gamma_n}{2}+\Gamma_i^*+\Gamma_n^*\right) \rho_{in}^{(1)} \label{eq:OBE5_4}\\
        \dot{\rho}_{nf}^{(2)} &=-i\Delta_b\rho_{nf}^{(2)}+\dfrac{i}{2}p_a^* \rho_{if}^{(3)}+\dfrac{i}{2}p_b(\rho_{ff}-\rho_{nn})-\left( \dfrac{\Gamma_n}{2}+\Gamma_n^*+\Gamma_f^*\right) \rho_{nf}^{(2)} \label{eq:OBE5_5}\\
        \dot{\rho}_{if}^{(3)} &=-i(\Delta_a+\Delta_b)\rho_{if}^{(3)}+\dfrac{i}{2}p_a^* \rho_{nf}^{(2)}+\dfrac{i}{2}p_b\rho_{in}^{(1)}-\left(\Gamma_i^*+\Gamma_f^*\right) \rho_{if}^{(3)} \label{eq:OBE5_6}\\
        \color{red}\dot{\rho}_{ss} & \color{red}=+\mathfrak{Im} \left( p_b^* \rho_{sg}^{(2)} \right) + \Gamma_n \rho_{nn} - \Gamma_n \mathfrak{B} \rho_{ss} \label{eq:OBE5_7}\\
        {\color{red}\dot{\rho}_{gg}} & {\color{red}= - \mathfrak{Im} \left( p_b^* \rho_{sg}^{(2)} \right)} \label{eq:OBE5_8}\\
        {\color{red}\dot{\rho}_{sg}^{(2)}} &{\color{red}=-i\Delta_b\rho_{sg}^{(2)}+\dfrac{i}{2}p_b(\rho_{gg}-\rho_{ss})-\left( \Gamma_s^*+\Gamma_g^*\right) \rho_{sg}^{(2)}} \label{eq:OBE5_9}
\end{align}
Here, we define $\rho_{sg}^{(2)}=e^{-i\omega_b t}\rho_{sg}$ . These equations constitute the model used for the numerical fit of the data.

\paragraph{Assumptions}

In order to simplify the fitting procedure, we assume that the initial state dephasing time is equal to the excited states \cite{hertel1996ultrafast} , i.e., $\Gamma_i^*=\Gamma_n^*=\Gamma_s^*$, and that the final states dephasing time is infinite ($\Gamma_f^*=\Gamma_g^*=0$). In this way, the only fitting parameters of the Gamma matrix are the inverse lifetime $\Gamma_n$ and the pure dephasing rate $\Gamma_n^*$ of the bright excitonic state at K,  together with the backscattering fraction $\mathfrak{B}$. 
However, a large number of parameters relative to the optical excitation require further optimization, in particular the determination of the time of coincidence of the pump and probe pulses. This quantity must be defined with a precision much higher than the measurable cross-correlation FWHM in order to achieve good confidence in the fit and determine lifetimes shorter than the duration of the pulses. 

Given the aforementioned approximations, reducing the band structure to a system of non-dispersing levels, a ploy may be used to achieve this. As mentioned in the main text, the density matrix treatment of the photoemission process upon resonant pumping allows correctly treating the mixing of the two types of quantum processes to reach the final state: the first one consists of a coherent two-photon process with an intermediate virtual state, and the second one involves a two-step process, where the A exciton is created at the K valley and then photoemitted to the final state by the probe photon.
The former can be separated from the latter by wavelength detuning of the pump to values smaller than the band gap ($\Delta_a < 0$). If the detuning becomes very large ($\Delta_a << 0$) the intermediate states of the two processes are very far apart in energy and they can be observed separately by solving the dynamics at two different $\Delta_b$. At probe detuning $\Delta_b=0$ the dynamics is only determined by the two-step process involving the formation of a real population in the K valley; while at probe detuning $\Delta_b=-\Delta_a$ the final state population arises from the coherent two-photon process with a detuned virtual state. The evolution in time of the latter allows to precisely assess the coincidence between the pump and probe pulses. 

To achieve a good fit, it is therefore advantageous to evaluate $\rho_{ff}$ once with a very large detuning (which we may define as $\Delta_a^{'}$) and extract its time evolution at $\Delta_b=-\Delta_a^{'}$ to isolate the coherent process dynamics and accurately determine the pump-probe coincidence. 
We can further evaluate $\rho_{ff}$ a second time at a pump detuning equal to the experimental pump detuning (which we define as $\Delta_a$), and extract its time evolution at $\Delta_b=0$ to determine the population dynamics of the K-exciton state. A third curve, arising from $\rho_{gg}$ evaluated at probe detuning $\Delta_b=0$ and pump detuning  $\Delta_a$, completes the set of three curves that can be extracted from the model and fitted to the experimental data, with the result plotted in Fig.3(a) of the main text. The effective coincidence time is the fourth (and last) fitting parameter.

\paragraph{Data selection}

In our data, the Floquet virtual intermediate state is clearly observable as a transient replica of the valence band shifted by 1.55 eV. The pump is slightly wavelength detuned from the direct optical bandgap at 1.67 eV, but this energy distance is not sufficient to clearly isolate the coherent process dynamics.
It is however possible to extract the time trace at very large detuning by integrating the ARPES dataset in a momentum range away from the K points (resonant transition) and in an energy window following the Floquet replica. This creates a fictitious large energy gap detuning, because the pump wavelength stays constant, but the energy gap for direct optical transitions increases as the conduction and valence band have diverging dispersion.

For the OBE fit reported in the main text, we select $\mathbf{k}_1= 0.84-0.96$ $\mathrm{\AA}^{-1}$, $\mathbf{E}_1= 1.11-1.26$ eV. Combining a fit of the conduction band minimum (see next paragraph) with theoretical calculations, we can estimate the effective energy gap for direct optical transitions at this point as $2.47\pm0.1$ eV, thus determining $\Delta_a^{'}=0.92\pm0.1$ eV. 
The energy range was chosen to be comparable to the energy resolution, while the momentum range was selected to include the full momentum distribution in the selected energy range. Small modifications of the momentum and energy windows did not affect the fit results.

Next, we extract the curve describing the K-exciton population dynamics. The EDC at +65 fs is fitted to locate the minimum of the conduction band, at $1.67\pm0.03$ eV; this also allows quantifying the effective $\Delta_a=0.12\pm0.03$. Then the momentum distribution curve (MDC) is fitted to find the centre of the valley. The curve is extracted in the range $\mathbf{k}_2=1.14-1.26$ $\mathrm{\AA}^{-1}$, $\mathbf{E}_2=1.57-1.6$ eV. Considerations on the energy and momentum ranges are the same as the previous paragraph.
The procedure to extract the position of the $\mathrm{\Sigma}$ valley is the same as for K, but in this case, since the $s$ level is taken to act as a sink for $n$ population, we need to account for all the electron scattered from K. Since in this case the carriers do have excess energy (the $\mathrm{\Sigma}$ valley is at lower energy than K), to eliminate the influence of intravalley dynamics we use a rather wide momentum-energy range: $\mathbf{k}_3=0.40-0.80$ $\mathrm{\AA}^{-1}$, $\mathbf{E}_3=1.17-2.17$ eV.  For a schematic of the method employed to extract the fitted signal, see Fig.~\ref{figS1_1}(b).

\paragraph{Parameter optimization and fit reliability}

\begin{sidewaysfigure}
\thispagestyle{empty}
\begin{center}
 \includegraphics[width=21 cm,keepaspectratio=true]{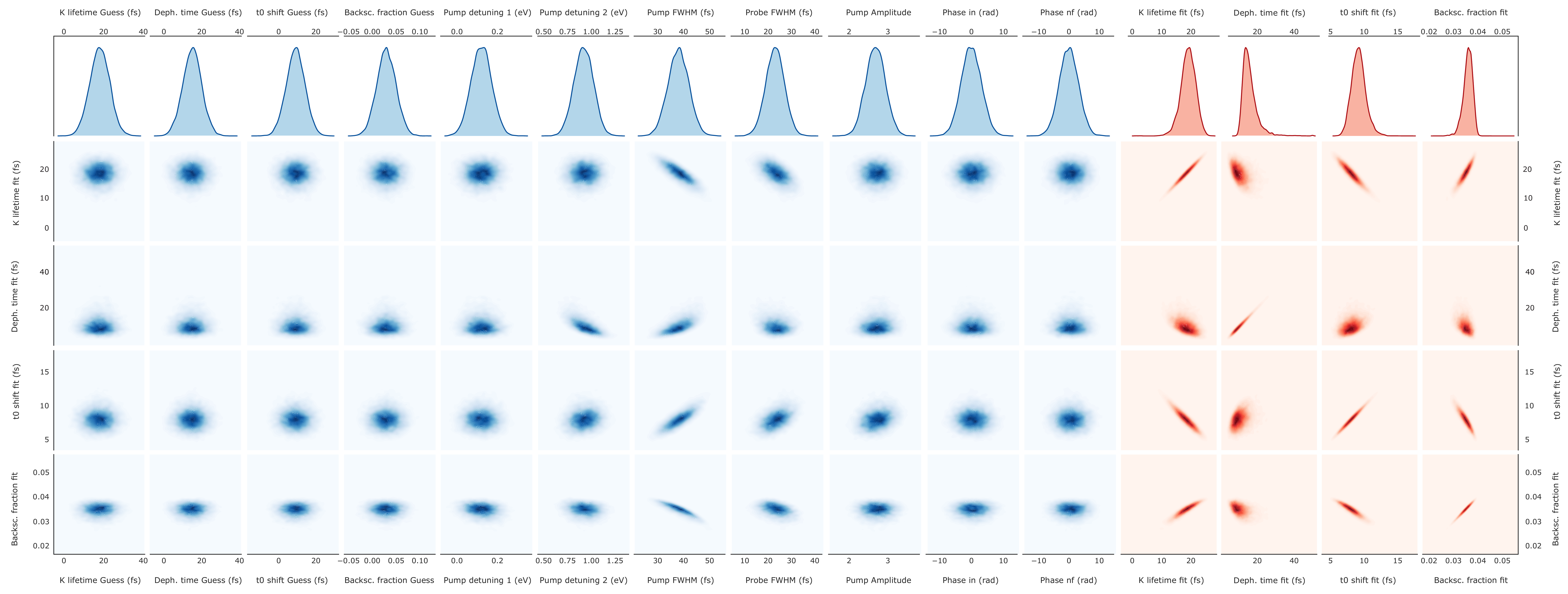}
 \caption{\footnotesize{\textbf{statistical analysis}. 
 Scatter plot matrix reporting the results of five thousand fits with normally distributed parameters around the estimated average and with the estimated standard deviations. The top line of graphs shows the univariate distribution of each input variable (blue), and output fit result (red). The lower four rows show density plots of bivariate distributions. The blue panels plot the distribution of the fit results against each input variable, while the red panels show the cross correlations between the fit results. Further details in the text.}}
 \label{figS1_2}
\end{center}
\end{sidewaysfigure}

The procedure to extract the data has allowed to use experimental results to fix two more parameters, the pump detunings $\Delta_a$ and $\Delta_a^{'}$. Only few parameters of the excitation pulses are left to be determined: the amplitudes of the electric fields, the time duration of each of the pulses, and the phase of the pump and probe fields.
The ratio between the amplitude of the fields can be reconstructed using the measured pulse energy of the pump and the photon flux of the probe and results $E_{0}^{pump}/E_0^{probe}=26000 \pm 2500$. As very large values of $E_{0}^{pump}$ create significant problems in the numerical evaluation of the OBE system, and anyhow we fit a normalized set of curves where only the ratio between the pump and probe signals is relevant, we set the fields to very small absolute values $E_0^{pump}=2.6 \pm 0.25$ V/m and $E_0^{probe}=0.0001 $ V/m. 
The choice of the phase is relevant for coherent spectroscopy applications such as interferometric photoemission or quantum beat spectroscopy: in these cases oscillations of the population are observable and their behaviour depends on the phases of the pump and probe fields \cite{weinelt2002time}. In our case they are not relevant, so we set them to zero. By repeating the fit for different choices of the phases, we indeed found no significant variation of the results (see correlations in Fig.~\ref{figS1_2}).

The pulse durations, in the form of the temporal FWHM of the Gaussian envelope, are critical parameters for the determination of lifetime and dephasing time. According to the results presented in \cite{puppin2019time}, we have rather accurate estimates for our set-up: $\mathrm{FWHM}_{pump}=32 \pm 2$ fs and $\mathrm{FWHM}_{probe}=19 \pm 2$ fs. However these values may show slight variations depending on the accuracy of the optimization procedure of the set-up.
To further refine these values for the conditions present during the specific experiment analyzed in this manuscript, we iterated the fit multiple times, fixing these two parameters at different values, and recorded the variation of the calculated uncertainty on the K-state lifetime and decoherence as shown in Fig.~\ref{figS1_1}(c-d). 

The errors reach minima at different values both of $\mathrm{FWHM}_{pump}$ and $\mathrm{FWHM}_{probe}$, but are well behaved in the range between the two minima, so we used the average and dispersion: $\mathrm{FWHM}_{pump}=36\pm3$ fs and $\mathrm{FWHM}_{probe}=23\pm2$ fs. The mean value of this optimization procedure have a discrepancy of about $4$ fs with respect to the values measured independently in \cite{puppin2019time}, so we adopted more realistic confidence intervals: $\mathrm{FWHM}_{pump}=36\pm4$ fs and $\mathrm{FWHM}_{probe}=23\pm4$ fs.
An analogous procedure has been used to confirm the values of the two detunings $\Delta_a$ and $\Delta_a^{'}$. In this case, minima of the errors were not observed. Rather, the error on the decoherence time decays exponentially, while the lifetime error varies slowly. However, the detunings at which the decoherence error drops to values comparable to the lifetime error are consistent with the experimentally determined $\Delta_a$ and $\Delta_a^{'}$, so we used those results.

Finally, we iterated the fitting procedure multiple times in order to determine realistically the confidence intervals. We repeated the fit while randomly choosing the values of the parameters within the assigned error bars (see Supplementary Tab.~\ref{tab:params}). 

\begin{table}[]
    \centering
    \begin{tabular}{c|c|c|c}
      Name & Value & Std. Dev. & Units \\
      Pump Detuning 1 & 0.122 & 0.05 & eV \\
      Pump Detuning 2 & 0.92 & 0.1 & eV \\
      Pump FWHM & 36 & 4 & eV \\
      Probe FWHM & 23 & 4 & eV \\
      Phase in & 0 & $\pi$ & rad\\
      Phase nf & 0 & $\pi$ & rad \\
      Pump amplitude & 2.6816 & 0.25 & - \\
    \end{tabular}
    \caption{\textbf{Fit parameters}. Fixed fit input parameters with their standard deviations as used in the optical Bloch equation fit.}
    \label{tab:params}
\end{table}

The latter procedure also allows observing the cross correlation between parameters and results of the fit, as displayed in Fig.~\ref{figS1_2}. Here, the univariate distributions of the input parameters (blue) and fit results (red) are plotted in the first row, showing that sufficient sampling is achieved to reach quasi-normal distributions in the inputs.
In the lower rows, the bivariate distributions display the distribution of each fitting result versus the input parameters (blue panels) or versus other fitting parameters (red panels). When the distributions have a round shape, they suggest statistical independence between the parameters. As an example, it is possible to look at the columns related to the phase parameters: the horizontal mirror symmetry of the distributions suggests that the results are independent from the phase choice. When the bivariate distributions assume a linear, diagonal shape, they indicate a strong interdependence (correlation) of fit parameters. For example, note how strongly the results are affected by the choice of pump FWHM, justifying the careful procedures used to determine its value.

To estimate the confidence interval, we calculated the standard deviation obtained from five thousand successive fits with parameters extracted randomly with normally distributed probability (see Supplementary Tab. \ref{tab:params} for mean values and standard deviation of each randomized parameter). We then estimated the 95\% confidence interval using a Student T distribution with 94 degrees of freedom (105 data points with 11 fixed parameters). The results are presented in Supplementary Tab.~\ref{tab:results}.

\begin{table}[]
    \centering
    \begin{tabular}{c|c|c|c|c}
      Name & Value & Std. Dev. & 95\% C.I. & Units \\
      $\tau_p$ & 18 & 2 & 4 & fs \\
      $\tau_\phi$ & 17 & 5 & 9 & fs \\
      $\Delta t_0$ & 9 & 1 & 2 & fs \\
      $\mathfrak{B}$ & 0.036 & 0.002 & 0.004 & - 
    \end{tabular}
    \caption{\textbf{Fit results}. Results of the optical Bloch equation fit, with the calculated error bars.}
    \label{tab:results}
\end{table}

\noindent
\ul{\textbf{Self-energy calculation}}

\noindent
For the density functional electron-phonon calculation we used the EPW code\cite{noffsinger2010epw} as part of the quantum espresso package\cite{giannozzi2009quantum} using norm-conserving pseudopotentials and the local density approximation. The hexagonal structure of WSe$_2$ was relaxed with lattice constants of a=6.2020 a.u, c=3.9548 a.u. and the density was computed with the pw.x program with a planewave cutoff of 160 Ry and a Monkhorst-Pack-grid of 42x42x16 k-points. The phonons were evaluated using the ph.x program on a q-mesh of 6x6x2. Based on a wannierization of the electronic structure using the wannier90.x program\cite{mostofi2014updated}, the EPW code interpolates the phonon and electron-phonon coupling quantities to a 20x20x10 q-mesh to evaluate the self energy, shown in the main text. 

The electron-phonon lifetime is evaluated at each band $n$ and k-point $k$ as, 
\begin{equation}
\begin{split}
 \frac{1}{\tau_{nk}} = \frac{1}{\hbar}\sum\limits_{m\nu}&\int\frac{dq}{\omega_{BZ}}|g_{nm\nu}(k,q)|^2\\
&\times[(1-f_{mk+q}+n_{q\nu})\delta(\epsilon_{nk}-\hbar\omega_{q\nu}-\epsilon_{mk+q})\\
&+(f_{mk+q}+n_{q\nu})\delta(\epsilon_{nk}+\hbar\omega_{q\nu}-\epsilon_{mk+q}))]     
\end{split}
\label{eq3}
\end{equation}
where the sum runs over all phonon modes ($q,\nu$) and electronic bands $\epsilon$. $g_{nm\nu}$ are the electron-phonon coupling matrix elements and $f$ and $n$ are the Fermi and Bose distributions, respectively.

\noindent
\ul{\textbf{Microscopic calculation of exciton signatures in trARPES}}

\noindent
The microscopic description of the time- and angle-resolved photoemission spectroscopy including the Coulomb interaction between electrons and holes and the exciton-phonon scattering dynamics is based on a many-particle Hamiltonian and the Heisenberg equation of motion formalism. We treat the quasi-particle band structure within a parabolic approximation, with effective masses stemming from \textit{ab initio} calculation established in literature \cite{kormanyos2015k}, and include the electron-light interaction for the VIS pump and the XUV probe pulse semi-classically in length gauge. In order to take into account the electron-hole Coulomb interaction we exploit the unit operator method exploiting the completeness relation of the Fock space\cite{ivanov1993self,katsch2018theory}. Hence, the conduction band electron operators are expressed uniquely by electron-hole pair excitations, which are transformed to the exciton picture by introducing exciton relative and center-of-mass momentum. The eigenenergies and wave functions of the excitons are computed by numerically solving the Wannier equation.
\begin{align}
\frac{\hbar^2\kp^2}{2m^{\xi_v\xi_c}}\varphi^{\xi_v\xi_c}_{\mu,\kp}-\sum_{\qp}V^{\xi_v\xi_c}_{\kp+\qp}\varphi^{\xi_v\xi_c}_{\mu,\kp}=E^{\xi_v\xi_c}_{\mu}\varphi^{\xi_v\xi_c}_{\mu,\kp}
\end{align}
with the index $\xi_{v/c}$ standing for the valley of the involved valence or conduction band electron, the reduced mass $m^{\xi_v\xi_c}$ and the exciton state $\mu$. The screening of the Coulomb potential $V_{\qp}$, due to the dielectric environment, is treated beyond the Rytova-Keldysh framework \cite{rytova1967the8248,keldysh1979coulomb, trolle2017model}.
To model the bulk we assume that the XUV pulse irradiates only the first two layers of the crystal. According to this, we take a bilayer band structure from \textit{ab initio} calculations \cite{roldan2014electronic} and solve the Wannier equation for the bilayer WSe$_2$ on a WSe$_2$ substrate. We find a value of \unit[91.3]{meV} consistent with the experimental result. 

The trARPES experiment measures a current of photoemitted electrons that the evaluated signal $I_{\mathbf{k},\varepsilon^f_{\mathbf{k}}}(\tau)=\lim_{t\rightarrow\infty}\int_{-\infty}^t dt' ~ \rho^f_{\mathbf{k}}(t',\tau)$ is determined by the vacuum electron distribution $\rho^f_{\mathbf{k}}=\langle f^{\dagger}_{\mathbf{k}}f^{\mathstrut}_{\mathbf{k}}\rangle$. The fermionic operator $f^{(\dagger)}_{\mathbf{k}}$ annihilates (creates) an electron in the continuum states of the vacuum. The signal is a function of in-plane momentum $\kp$ and kinetic energy $\varepsilon^f_{\mathbf{k}}$ of the photoelectrons and pump-probe time delay $\tau$. Note, that the wave vector $\mathbf{k}$ is three-dimensional with its in-plane component $\kp$. The equation of motion for the vacuum electron population in exciton basis reads \cite{christiansen2019theory}
\begin{align}
\frac{d}{dt}\rho^f_{\mathbf{k}}=-2\im \left( \Omega^{vf}_{\mathbf{k}} P^{vf}_{\kp,\mathbf{k}} + \Omega^{cf\xi_c-}_{\mathbf{k}} P_0^{*\xi_c\xi_v}P^{vf\xi_v}_{\kp,\mathbf{k}} + \sum_{\xi_v,\mathbf{Q}} \Omega^{cf\xi_c}_{\mathbf{k,Q}} \delta\langle P_{\mathbf{Q}}^{*\xi_c\xi_v}P^{vf\xi_v}_{\kp-\mathbf{Q},\mathbf{k}}\rangle \right)
\end{align}
restricted to the 1s-A-exciton. The first term accounts for photoemission of valence band electrons $P^{vf}_{\kp,\mathbf{k}}=\langle v^{\dagger\xi_v}_{\kp}f^{\mathstrut}_{\mathbf{k}}\rangle$ with the corresponding Rabi-frequency $\Omega^{vf}_{\mathbf{k}}$. The second and third terms describe the exciton-assisted photoemission yielding the excitonic signals in trARPES, with the corresponding Rabi-frequencies $\Omega^{cf\xi_c-}_{\mathbf{k}}=\Omega^{cf\xi_c}_{\mathbf{k}}\varphi_{\kp}^{*\xi_c\xi_v}\exp(i\omega_{vis}t)$ and $\Omega^{cf\xi_c}_{\mathbf{k,Q}}=\Omega^{cf\xi_c}_{\mathbf{k}}\varphi_{\kp-\alpha^{\xi_c}_{\xi_v}\mathbf{Q}}^{*\xi_c\xi_v}$. The exciton wavefunction is denoted by $\varphi_{\kp-\alpha^{\xi_c}_{\xi_v}\mathbf{Q}}^{*\xi_c\xi_v}$ with center-of-mass momentum $\mathbf{Q}$ and mass ratio $\alpha^{\xi_c}_{\xi_v}=m_e/(m_e+m_h)$. The second term couples the excitonic coherence $P_0^{*\xi_c\xi_v}$, excited by the VIS pump pulse, with the photoemission transition of valence band electrons. In contrast, the third term is driven by the excitonic population caused by the phonon-induced decay of the excitonic coherence \cite{selig2016excitonic,christiansen2017phonon,raja2018enhancement}.We find a $T_2^*$ time of \unit[18.1]{fs}. The exciton dynamics describe the phonon-induced relaxation and thermalization throughout the Brillouin zone expressed by a Boltzmann-like scattering equation \cite{selig2018dark}. For the exciton-phonon interaction the underlying electron-phonon deformation potential matrix elements for acoustic and optical phonons are taken from \textit{ab initio} calculations \cite{jin2014intrinsic}. In a first approximation, we use for the bilayer the same electron-phonon matrix elements as for the monolayer, which is supported by similar temperature dependent shifts in absorption experiments when going from mono- to bilayer \cite{raja2018enhancement}. The trARPES signal for small pump-probe delays is determined by the second term. The excitonic coherence oscillates with the excitation energy and decays with increasing time forming incoherent excitons, lying at the exciton energy, which explains an observable time dependent shift of the excitonic trARPES signal for detuned excitations. At large pump-probe delays the third term is the origin of the excitonic trARPES signal.

The slight deviation of the MDC between theory and experiment might be explainable by the spectrally broad pump pulse used in the experiment, which generates a hot exciton distribution of KK excitons (electron and hole situated at the K-point), which broadens the momentum distribution curve even at $\tau = 0$.



\bibliographystyle{cite}

\clearpage